# Comment: Microarrays, Empirical Bayes and the Two-Groups Model

**Carl N. Morris**

*Abstract.* Brad Efron's paper has inspired a return to the ideas behind Bayes, frequency and empirical Bayes. The latter preferably would not be limited to exchangeable models for the data and hyperparameters. Parallels are revealed between microarray analyses and profiling of hospitals, with advances suggesting more decision modeling for gene identification also. Then good multilevel and empirical Bayes models for random effects should be sought when regression toward the mean is anticipated.

## 1. FREQUENCY, BAYES, EMPIRICAL BAYES AND A GENERAL MODEL

Brad Efron's two-groups approach and the empirical null ("null" refers to a distribution, not to a hypothesis) extension of his local fdr addresses testing many hypotheses simultaneously, with modeling enabled by the repeated presence of many similar problems. He assumes two-level models for random effects, developing theory by drawing on and combining ideas from frequency, Bayesian and empirical Bayesian perspectives. The last half-century in statistics has seen exciting developments from many perspectives for simultaneous estimation of random effects, but there has been little explicit parallel work on the complementary problem of hypothesis testing. That changes in Brad's paper, especially for testing many hypotheses when exchangeability restrictions are plausible.

"Empirical Bayes" is in the paper's title, said in Section 3 to be a "bipolar" methodology that draws on frequency and Bayes, but otherwise with a meaning left for us to infer from the paper's example datasets. The examples all involve two-level models with inferences about many unknown parameters, that is, about the unknown random effects. Blending frequency and Bayesian thinking in statistics will be appreciated especially by statisticians who engage both in theoretical and in applied research, and we know that many of statistics' best and time-honored procedures perform well simultaneously from the frequency and the Bayesian perspectives. Classifying statisticians as either Bayesian or frequentist ignores the fact that these terms have varying meanings to different statisticians, and it encourages the view that statisticians must adopt just one of these perspectives exclusively, which many statisticians, myself included, do not do.

*Carl N. Morris is Professor, Department of Statistics, Harvard University, Cambridge, Massachusetts 02138, USA (e-mail: morris@stat.harvard.edu).*

The frequency perspective requires comparing procedures on the basis of repeated sampling, but it can be neutral about how procedures are constructed. The Bayesian approach, after a model is completely specified, including the "prior" ("structural" or "mixing" might be better adjectives) distribution, must use the laws of probability to assess uncertainties about unknowns, given the observed data and the model. Valuably even from the frequency perspective, Bayesian reasoning can be used to suggest how to construct inferences about population parameters and other unobservables, at least in ideal settings. That is illustrated in Efron's treatment of the fdr and the Fdr here. Such modeling of likelihood functions at more than one level, and of priors, becomes less subjective when one has more data, especially with massive datasets like those in the paper.

Scientists have been encountering ever more massive datasets, especially since modern censoring technology and computers have evolved to make collecting, organizing, visualizing and analyzing such databases possible. My early experience in the 1980s involving two-level models for NASA's satellite imagery data made it clear to me that science had reached a new point where computers not only had enabled us to analyze large datasets, they even made it possible to collect very large datasets. The computer had become a horse that could collect and analyze data as we directed, and statisticians were its jockeys. While the massive datasets we see today can be overwhelming, Brad rightly recognizes that they can be welcomed as opportunities to build better models. That not only leads to more accurate inferences for the given data, but better models also advance knowledge and future scientific discoveries.

Brad's use of "empirical Bayes" with the six datasets of the paper is restricted to datasets he considers to be exchangeable. That could signify his moving away from a liberalizing view of empirical Bayes that we once developed together. I doubt this, but the analyses shown assume that the joint distribution of the data and of the random effects are exchangeable. Our papers together in the 1970s moved empirical Bayes away from that requirement, partly to provide a perspective from which acceptable shrinkage generalizations of Stein's estimator might be developed. That was and is needed especially when (nearly) unbiased estimates of the different random effects have different variances, perhaps most often because of different sample sizes.

In seeking a firmer basis for modeling and inference in empirical Bayes settings (Morris, 1983), I continued back then to use that term. However, Herb Robbins, who coined the term, made it clear to me then that the version he had pioneered, built around exchangeability, asymptotics and nonparametric mixing (prior) distributions, was how he wanted the term to be used. Also about then, D. V. Lindley averred that "There is no one less Bayesian than an empirical Bayesian," a comment that seemed mainly directed at Robbins' approach. Some other statisticians then, and perhaps still today, thought of empirical Bayes as restricted to plugging hyperparameter estimates into Bayes rules. So the term "empirical Bayes" meant different things to different statisticians, and not always good things.

It also had become clear to me back then that dealing with many inferences simultaneously had to be guided by Bayesian reasoning. For example, Bayesian constructions show why interval estimates based on plug-in methods can be much too narrow, especially when the number ($N$, in the notation of Brad's paper) of random effects being estimated is small or moderate. So I began to use the term empirical Bayes more sparingly to describe my own work. In building on the ideas behind my 1983 paper, and when trying to combine frequency ideas with Bayes in hierarchical models, I sometimes have referred to a "general model for statistics" for the desired frequency/Bayes unification.

The general model includes distributions for data given parameters of interest, and for the hyperparameters that govern the distribution of those parameters, conceptually (but not always) specified for at least two hierarchical levels. From the frequency perspective in this general model, all possible distributions would be considered for the hyperparameters, those being mixtures of atomic (Dirac) distributions. From the subjective Bayesian perspective, just one distribution (a prior at the top level of the hierarchy) would be allowed in a particular inferential problem. (This framework extends to nonparametric models by letting the parameters and/or hyperparameters be infinite dimensional.)

This general model puts frequency and Bayesian models at the endpoints of a continuum, with the middle span open for flexibly specifying restrictions on distributions that could accommodate empirical Bayes and other models. Decision theory extends to this general model so that frequency (resampling) evaluations would be done conditionally for the range of hyperparameters. Such resampling was carried out when evaluating the coverage probabilities of parametric empirical Bayes interval estimates in Morris (1983) and in much other work since then. In a University of Texas dissertation, Joe Hill showed how this general framework extended to ancillarity, information, and other fundamental statistical ideas (Hill, 1986, 1990).

Aside from their different interpretations, the frequency and Bayesian perspectives can be quite complementary. The frequency paradigm is normative, but not necessarily prescriptive. The fundamental theorem of (frequency) decision theory, that is, the complete class theorem, supports the Bayesian connection by recognizing that the admissible procedures nearly coincide with the class of extended Bayes rules. Statistical procedures with good repeated sampling (frequency) properties often can be anticipated by thinking about Bayesian constructions.

A reminder of how Bayesian procedures can have better frequency properties than those derived solely by frequency reasoning is illustrated by a graph with $N=15$ in Christiansen and Morris (1997, Figure 1). Poissonly distributed summary data like those seen at heart transplant hospitals are fitted there via two-level models. The graph there shows the coverage rates in repeated sampling of nominal 95% intervals when the transplant success rates are simultaneously estimated at the different hospitals. Six procedures are evaluated. Two follow Bayesian constructions, one that uses the BUGS program and default prior, and the other being an accurate approximation of a hierarchical Bayes procedure based on a hyperparameter prior akin to Stein's superharmonic prior for Normal distributions. These two Bayesianly motivated interval procedures cover or nearly cover 95% of the time in repeated sampling simulations, as intended. The four frequency procedures based on MLE, REML and on two GLM multilevel techniques, have coverages in the range of 60% to 90%, falling well below the claimed coverage rate of 95%. Whether developed from Bayesian or frequency considerations, good frequency procedures must provide coverages in repeated sampling close to their claimed values, but the four non-Bayesian procedures do not meet that standard.

## 2. FDR, FDR AND EXCHANGEABILITY

Brad illustrates the use of Bayesian modeling and probabilistic reasoning with his six large datasets to produce approaches to hypothesis testing that would be valid if prior information were available. Then he shows how to estimate the needed prior, or mixing, distributions from repeated data.

Probabilistic modeling leads directly to Efron's local fdr, which in turn leads to the Benjamini–Hochberg Fdr procedure. Starting with the simplest "two-groups" model, with density $f_0$ under the null hypothesis $H_0$ and $f_1$ under the alternative hypothesis $H_1$, the paper moves through increasingly elaborate probability models discovered in the process of modeling and analyzing exchangeable data and repeated problems. Benjamini and Hochberg's celebrated false discovery rate statistic Fdr applies when all the $H_0$ distributions have a single theoretically determined density function $f_0$, and when the prior probability $p_0$ of $H_0$ is high (at least 0.9). Then $f_1$, the $H_1$ density, is available via estimating the marginal density, $f(z) = p_0 * f_0(z) + p_1 * f_1(z)$ and solving for $f_1(z)$. While $f_1$ is not actually needed in exchangeable cases, it will be for a nonexchangeable extension which I will review later. Thus, a direct estimate of the posterior probability of $H_0$, given the data, only requires $p_0$, $f_0$ and $f(z)$ in this simplest case.

This approach is beguilingly simple, but its validity depends crucially on a restrictive exchangeability assumption that can be missed. The marginal density $f(z)$ will be the same for all the $z_i$ observations only if the same $f_1$ distribution holds under $H_1$ for all $z_i, i=1,\ldots,N$. This may hold for five of the six datasets in the paper, but it does not for the school data, as discussed later.

As formula (2.7) shows, the local fdr is the posterior probability of $H_0$, that is,

$$\mathrm{fdr}(z) = P(H_0|Z=z) = \frac{p_0 * f_0(z)}{p_0 * f_0(z) + p_1 * f_1(z)}.$$

Starting with $\mathrm{fdr}(z)$ before introducing $\mathrm{Fdr}(z)$ seems natural, but this particular history has developed oppositely. Efron's local fdr is immediately interpretable in probabilistic or Bayesian terms because choosing between hypotheses $H_0$ and $H_1$ means considering $P(H_0|z)$, and also because fdr depends on the likelihood ratio, and on the Neyman–Pearson statistic.

As Brad writes, the Benjamini–Hochberg Fdr statistic (2.3) is the integral of $\mathrm{fdr}(z)$. Starting with $\mathrm{fdr}(Z) = P(H_0|Z)$ and assuming that one will choose $H_1$ whenever $Z \leq z$ leads to

$$E(\mathrm{fdr}(Z)|Z \leq z) = P(H_0|Z \leq z) = \mathrm{Fdr}(z),$$

as shown in the paper, and this is

$$\mathrm{Fdr}(z) = \frac{p_0 * F_0(z)}{p_0 * F_0(z) + p_1 * F_1(z)}.$$

Thus, $\mathrm{Fdr}(z) = P(H_0|Z \leq z)$ is the fraction of times that $H_0$ would be falsely rejected. The Benjamini–Hochberg false discovery rate $\mathrm{Fdr}(z)$ is discovered probabilistically as the average probability (the pre-posterior probability in Bayesian terms) of accepting, that is, discovering, $H_1$ falsely.

The probability model that leads to the fdr and Fdr statistics in repeated applications assumes exchangeability in two ways. First, $p_0$ should not depend on $i$, as Efron discusses in Section 2. Second, $f_0$ and $f_1$ must be the same for all problems $i = 1, 2, \ldots, N$. From the two-level modeling perspective of the paper, $f_1(z)$ is a mixture of densities for the (approximately) $N * p_1$ values of $\mu_i$ that are distributed according to $H_1$. Denoting the random effects as $\mu_i$ for $i = 1, \ldots, N$, exchangeability permits the conditional densities $f(z_i|\mu_i)$ for $z_i$ to depend on $i$ through $\mu_i$ only, and not otherwise to depend on $i$.

Some two-level settings are modeled with "paired" exchangeability among individuals [i.e., the collection of pairs $(z_i, \mu_i)$ are exchangeable], and that produces exchangeability for the marginal distributions of $z_i$. This happens familiarly with $N$ independent individuals (in the paper, "individuals" are the $N$ genes, and the schools, etc.) if the joint distributions of $(z_i, \mu_i)$ are i.i.d. Robbins' original introduction of empirical Bayes for Poisson models rested on paired exchangeability because every individual Poisson distribution was assumed in his paper to have the same exposure. The James–Stein estimator arises as a parametric empirical Bayes estimator, but only when paired exchangeability holds, as when the sample means all have the same variances.

A happy consequence of pairwise exchangeability is that Bayesian procedures often can conveniently be expressed explicitly in terms of the marginal (unconditional) distribution of the data $(z_i)$, and that marginal can be estimated directly from the observed $z_i$, as Efron has done in several settings. This gives an asymptotically consistent estimate of a Bayes procedure, and the statistician then can avoid directly estimating the mixing distribution $g(\cdot)$ that governs the random effects, $\mu_i$. Relatively simple expressions then may emerge, such as the procedures of Robbins, of Stein, and of Benjamini–Hochberg. As Efron notes, the independence assumption is not crucial, but exchangeability is. The Fdr and fdr statistics in the exchangeable setting of Efron's Section 2 should work well with pairwise exchangeability when $N$ is large, but exchangeability can be restrictive and may depend heavily on prior knowledge. Seemingly, exchangeability is widely considered to hold for microarray, proteomics, BRCA and spectroscopy data. It cannot be valid for the school data because school enrollments, that is, sample sizes $n_i$ vary. Nearly all theory presented in this paper is based on such exchangeability, barring the discussion of nonexchangeable choices for $p_0$ in Section 2. Is "empirical Bayes" in this paper meant to be limited to exchangeable (or pairwise exchangeable) settings?

## 3. MULTIPLE HYPOTHESIS TESTING—LOOKING FOR LARGE RANDOM EFFECTS

Here is an extension of Efron's approach that may be especially useful for identifying large random effects $\mu_i$. First consider and fix any single value of $i$, $1 \leq i \leq N$, with $z = z_i$ having been observed, and assume that the "theoretical null" $N(0, 1)$ distribution holds for $z_i$ under $H_0$, that is, when the random effect $\mu = \mu_i = 0$. Assume $p_0$, $f_0$ and $f_1$ all are known for this value $i$, as in Section 2, and that $g(\cdot)$ is known. Then $f_1(z)$, the marginal distribution of $z$ under $H_1$, is determined by integrating the conditional distribution of $z$ given $\mu$, for example, $z \sim N(\mu, 1)$ having density $\phi(z - \mu)$, over the distribution $g(\mu)$ that governs the $H_1$ distribution of $\mu$. (Exchangeability does not matter when all these distributions are known.) Then when $H_1$ holds, the density of $\mu$ given $z$ is

$$h(\mu|z) = \phi(z - \mu) * g(\mu) / f_1(z).$$

With $\mathrm{fdr}(z) = P(\mu = 0|z)$, and writing $\delta(\mu)$ as the Dirac delta function ($\mu = 0$ with probability 1 when $H_0$ is true), the density of $\mu$ given $z$ is expressible as a mixture of Efron's $\mathrm{fdr}(z)$ according to

$$p(\mu|z) = \mathrm{fdr}(z) * \delta(\mu) + (1 - \mathrm{fdr}(z)) * h(\mu|z).$$

If all these distributions and values were known, one could "test" $H_0 : \mu = 0$ (or $\mu \leq 0$?) versus $H_1 : \mu > 0$ by using $\mathrm{fdr}(z)$ as the probability of $H_0$. However, one well might prefer only to identify genes "far from $H_0$," that is, only select values of $\mu > k$ that exceed a scientifically substantial magnitude $k > 0$, and with a substantial probability. One then would use $p(\mu|z)$ in the formula above to calculate $P(\mu \geq k|z)$.

Numerical illustrations are easy to do, and here is one based on the assumptions in Section 5 of the paper, with $N = 3000$, $p_0 = 0.9$, and Normal distributions with $z_i \sim N(\mu_i, 1)$ and $g(\mu_i)$ being the $N(2.5, 0.5)$ density. Then values of $z \geq 3.5$ occur in 2.1% of the genes, so $z \geq 3.5$ identifies about 63 of the 3000 genes. If we were to choose $k = 2.8$, then $P(\mu > 2.8|z) = 0.506$ at the threshold value $z = 3.5$, and the conditional probability that $\mu > 2.8$ rises as $z$ increases. Researchers who wish to identify about 63 genes (2.1%) would calculate $P(\mu_i > 2.8|z_i)$ for every one of the 63 selected genes, all those that have at least a 50% chance of $\mu > k = 2.8$, and (by averaging) that overall about 60% of the 63 selected cases have $\mu_i > 2.8$. The 60% statement is analogous to Benjamini–Hochberg's calculation, calculated here by averaging the 63 selected posterior probabilities. If a smaller value $k = 2.0$ were chosen, then selected genes at that threshold, still with $z \geq 3.5$, would have at least a 90% chance (90% if $z = 3.5$ exactly) that $\mu > 2.0$, and one would know that about 95%, or 60 of the 63 selected cases, would have $\mu > 2.0$. Of course, if $k = 0$, as in the paper, then fdr$(z)$ and $F(z)$ would indicate that about 98% (61 or 62) of the 63 selected cases with $z \geq 3.5$ would have $\mu > 0$.

The preceding assumes a one-tailed test, as does Fdr, and so we have used $k > 0$ (if large positive values of $\mu$ are wanted), but two-tailed probabilities also are easy to evaluate. A table of the $N = 3000$ genes could list genes, sorted by their values of $P(\mu > k|z)$, using $p(\mu|z)$. With exchangeability, the ordering is that of $z_i$. Researchers could review these values of $P(\mu > k|z)$, keeping as many genes as desired, and stop when this probability becomes too low, or when enough candidates have been accepted. There is nothing special about keeping 2.1% and changing the cutoff for $z$ would alter that percentage. Experience gained with different values of $k$ after a variety of analyses with various data sets eventually might help identify the scientifically most useful values.

Of course $g(\mu)$ and the other constants are not generally known. That is the point of Efron's paper, but $g$ can be estimated by a variety of methods, frequentist, Bayesian and empirical Bayesian, and perhaps quite accurately with large $N$. The paper shows some nifty ways to estimate $f_1$ in exchangeable settings. Then one could use the estimated $f_1$ to estimate $g(\mu)$, perhaps by deconvolution methods. While estimating these mixing distributions $g(\cdot)$ becomes more difficult in nonexchangeable cases when the $z_i$ have different conditional distributions given $\mu_i$, the literature provides a variety of ways to do that, most easily in parametric settings.

The proposal just described would test interval null hypotheses instead of single points by calculating $P(H_1)$ given the data, also by using the data to learn about various constants and distributions, for example, about $p_0, g(\cdot)$, etc. Doing this in conjunction with choosing a $k > 0$ has been recommended in medical profiling by Burgess, Christiansen, Michalak and Morris (2000) for profiling hospital performances. Standard practice for medical profiling most commonly is based on testing different hypotheses like $H_0 : \mu_i = 0$ independently, using standard methods like those widely taught in beginning statistics courses. That forfeits the possibility of developing more information via multilevel modeling. Once multilevel models have been fitted, it is natural to consider alternative hypotheses like $H_1 : \mu > k$ where $k > 0$ is chosen to set standards ($k$) for unacceptable (or laudatory) departures from average outcomes of medical procedures. The analogous proposal is made here, which can be extended to accommodate a spike at 0 with $p_0 > 0$ within $H_0 = (-k, k)$ if required. That extension is not needed with medical profiling data, where it is unlikely that any sizeable fraction $p_0$ of hospitals would have precisely the same underlying rates of surgical outcomes, but the paper's applications make it clear that positive probability for a null point within $H_0$ is appropriate in a variety of problems.

In exchangeable cases, ranking according to $p$-values will not depend on the choice of $k$. With medical outcome data for hospitals, the number of treated patients always will vary substantially, producing nonexchangeability. Then shrinkages toward a common mean will be greater for small hospitals than for large ones, and the resulting rankings will depend not only on $z_i$, but also on $n_i$ and on $k$.

## 4. NONEXCHANGEABILITY, THE SCHOOL DATA AND THE ONE-GROUP MODEL

The school data of Figure 1(b) are not exchangeable because the sample sizes $n_i$ (actually there are two different sample sizes for each school, one for each demographic group) surely vary across the $N = 3748$ schools. Equal sample sizes might lead to exchangeability, but that rarely happens except with designed experiments, as the microarray experiments must be. Together (e.g., in Efron and Morris, 1975), Brad and I once used toxoplasmosis summaries for $N = 36$ regions to illustrate generalizations of Stein's estimator that were needed to account for different sample sizes in different regions. Those toxoplasmosis data, the hospital profiling data, and the school data in this paper all might be similarly modeled. The school data calculations suggest shrinkages should vary, but average about 40%. A sharp null with $p_0$ much in excess of 0 seems implausible for toxoplasmosis, for hospital data, and for the school data, and so Efron introduces the case $p_0 = 0$ as his "one group model." One would then expect that Var$(z_i)$ is proportional to $n_i$. That would cause longer-tailed distributions for the $\{z_i\}$ values than Normality allows, and schools with more students would tend to be the outliers. Figure 1(b) reveals evidence of such long tails, corresponding to nonexchangeability.

## 5. INTERVAL ESTIMATION

Efron's Section 7, about interval estimation, shows in a simulation with exchangeable data that the FCR intervals are too wide. That happens because the FCR does not adjust its slope to be less than 1.0 when a gentler slope closer to 0.5 would track regression toward the mean (RTTM) of the 1000 random effects better. Interval estimates recentered according to this slope improvement can be shorter and still cover at the same rate as FCR does. Morris (1983) provides a basis for evaluating interval coverages via repeated sampling. Figure 1 in that 1983 paper (data for $N = 18$ baseball players from some early Efron–Morris papers) illustrates how intervals centered on shrunken estimates are much more accurate. The graph there makes the same point that Efron does in Figure 8. However, Brad's Section 7 conclusion avers that Bayesian intervals cannot be trusted. That does not square with my experience because I have found Bayesian reasoning to be essential to understanding how to construct interval estimates that have good frequency properties.

With 1000 observations it makes sense to estimate the distribution $g(\mu)$ without assuming Normality, and instead to use exchangeability as a basis for estimating the marginal distribution of the $\{z_i\}$. The same can be done with Bayesian methods, even with a nonparametric specification for $g$, although less easily. With unequal sample sizes, or when $N$ is not large, a Bayesian approach may be more successful, as with the heart transplant data of Christiansen and Morris (1997). A key to knowing that Bayesianly constructed confidence intervals will meet frequency resampling criteria requires identifying and using frequency-friendly noninformative distributions for the hyperparameters. This has been done in a variety of specific parametric settings, including for some common generalized linear models. Bayesian reasoning also shows us how to account for added variability in settings where the hyperparameters and shrinkage constants have been estimated. Such intervals must bow outward in Efron's Figure 8 when moving away from the center, and this is seen more dramatically when $N = 18$ in Figure 1 of Morris (1983). Efron's Figure 8 shows no bowing, but that would be too small to see with such large sample sizes. More discussion is needed as to whether Bayesian reasoning really has failed in the Section 7 setting, and about what an empirical Bayes approach really can offer, beyond suggesting Bayesian methods designed to withstand frequency verifications.

## 6. MODELING AND RTTM

Two-level modeling can reveal by how much random effects will regress (shrink) toward the mean (RTTM). The modeling task is to estimate the mean to shrink toward, and determine how much shrinkage. A term I always liked that Brad used when we wrote together is "ensemble information." RTTM means individual estimates will regress toward the ensemble estimates.

The paper focuses on rectangular $X$ as an $N$-by-$n$ data matrix. When $X$ is rectangular, it is especially valuable to analyze the distribution of the rows and columns of $X$, calculating correlations as Brad does among the rows (genes) and/or among the columns (arrays) to improve estimates of $f_0$, $f_1$ and $p_0$, and thereby to keep modeling assumptions at a minimum.

Of course $X$ need not be rectangular, nor should it automatically be so considered, because different rows sometimes may contain different amounts of data. The school data would follow such a nonrectangular shape if each row were to include separate entries for each student (as the BRCA and the HIV data do, but always with the same sample sizes). In this case, the school data have been forced into a rectangular Procrustean bed by using summarized data $z_i$, and that has obscured their nonexchangeability.

Sometimes it pays to take advantage of situations when $N$ is large, but without appealing to asymptotics. In the context of the paper, that might be done by increasing the number of parameters and fitting richer models as $N$ increases. This is parametric model-building, but it is an alternative to nonparametric modeling. The paper does some of this to investigate correlations, but the same could be done to assess whether exchangeable models are adequate.

A model for microarray data considered by Hongkai Ji and Wing Wong (Ji and Wong, 2005), concerns dealing with the (nuisance) standard deviations $\sigma_i$ that are estimated in the denominator of each $t$-statistic, like those considered in Sections 4 and 5 in Efron's paper. The sample standard deviations $s_i$ (each based on just a few degrees of freedom, as with the BRCA and HIV data) easily can produce randomly small sample standard deviations to estimate $\sigma_i$, and hence produce large $t$-values that falsely indicate which genes are expressing themselves importantly. A way out of this is to consider the $N$ problems to be exchangeable with respect to the random effects $\mu_i$ and also the $\sigma_i$. That justifies shrinkage methods (based on chi-squared distributions). Ji shows that shrinking the sample standard deviations $s_i$ toward their common mean, and using these empirical Bayes shrunken estimates in place of $s_i$ in the $t$-statistics, greatly improves the rate of false gene discoveries.

## 7. CONCLUSION

Brad Efron's paper introduces many ideas for analyzing massive datasets. It encourages a frequency-Bayes unification and empirical Bayes modeling. The paper identifies modeling and inference opportunities that arise with massive datasets in exchangeable settings. Much remains to do to understand the exchangeable case for parametric and nonparametric models alike, and there is much to do to recognize when nonexchangeable models are required, and how to fit them.